\documentclass[sigconf]{acmart}

\usepackage{courier}
\usepackage{multirow}
\usepackage{multicol}
\usepackage{xcolor}
\usepackage{stfloats}
\usepackage[breakable]{tcolorbox}

\usepackage{paralist}

\usepackage[draft=true]{minted}

\usepackage{listings}
\usepackage{xspace}
\usepackage{makecell}

\newcommand{\autoware}{Autoware\xspace}

\AtBeginDocument{%
  }

\setcopyright{acmlicensed}
\copyrightyear{2018}
\acmYear{2018}
\acmDOI{XXXXXXX.XXXXXXX}
\acmConference[Conference acronym 'XX]{Make sure to enter the correct conference title from your rights confirmation email}{June 03--05, 2018}{Woodstock, NY}
\acmBooktitle{Companion Proceedings of the 33rd ACM Symposium on the Foundations of Software Engineering (FSE '25), June 23--27, 2025, Trondheim, Norway}
\acmISBN{978-1-4503-XXXX-X/18/06}

\begin{document}

\title{Fine-grained Testing for Autonomous Driving Software: a Study on Autoware with LLM-driven unit testing}

\author{Wenhan Wang}
\authornote{Equal contribution}
\email{wenhan12@ualbera.ca}
\affiliation{%
  \institution{University of Alberta}
  \country{Canada}
}

\author{Xuan Xie}
\authornotemark[1]
\email{xxie9@ualbera.ca}
\affiliation{%
  \institution{University of Alberta}
  \country{Canada}
}

\author{Yuheng Huang}
\email{yuhenghuang42@g.ecc.u-tokyo.ac.jp}
\affiliation{%
  \institution{The University of Tokyo}
  \country{Japan}
}

\author{Renzhi Wang}
\email{renzhi.wang@ualberta.ca}
\affiliation{%
  \institution{University of Alberta}
  \country{Canada}
}

\author{An Ran Chen}
\email{anran6@ualberta.ca}
\affiliation{%
  \institution{University of Alberta}
  \country{Canada}
}

\author{Lei Ma}
\email{ma.lei@acm.org}
\affiliation{%
  \institution{The University of Tokyo}
  \country{Japan}
}
\affiliation{%
  \institution{University of Alberta}
  \country{Canada}
}

\renewcommand{\shortauthors}{}

\begin{abstract}

  Testing autonomous driving systems (ADS) is critical to ensuring their reliability and safety. Existing ADS testing works focuses on designing scenarios to evaluate system-level behaviors, while fine-grained testing of ADS source code has received comparatively little attention. To address this gap, we present the first study on testing, specifically unit testing, for ADS source code. Our study focuses on an industrial ADS framework, Autoware. We analyze both human-written test cases and those generated by large language models (LLMs). Our findings reveal that human-written test cases in Autoware exhibit limited test coverage, and significant challenges remain in applying LLM-generated tests for Autoware unit testing. To overcome these challenges, we propose AwTest-LLM, a novel approach to enhance test coverage and improve test case pass rates across Autoware packages.
\end{abstract}



\maketitle

\section{Introduction}

Autonomous Driving System (ADS) testing is pivotal for ensuring the safety and reliability of Autonomous Vehicles (AVs), and 
traditional ADS testing methods~\cite{tang2023survey, tian2018deeptest, zhang2018deeproad, cheng2023behavexplor} primarily focus on designing scenarios to evaluate the overall system behavior, with the main objective of verifying whether the ADS can make correct decisions to complete tasks with no dangerous behaviors. In contrast, more fine-grained ADS testing, especially testing on the source code implementation of different ADS modules, has not been carefully investigated. One of the major forms of fine-grained testing is unit testing \cite{lou2022testing}, which involves generating test cases to test individual functions in isolation. Unit testing is beneficial for ADS software as it enables developers to efficiently identify and address potential bugs at the function level.

In this paper, we conduct an exploratory study on source code-level unit testing for ADS.
Specifically, we perform study on Autoware~\cite{autoware}, which is a widely recognized ADS platform and developed by our industrial partner, TIER IV. 


As an industrial software system mainly written in C++, Autoware has more than 200K lines of code across 10 ADS modules, and many Autoware packages may depend on one or more other packages.
Autoware utilizes the Robot Operating System (ROS) \cite{ROS} as its core mechanism for message passing and synchronization between components. Its development toolchain (e.g., the \texttt{colcon} build system), node management (e.g., parameter server), and debugging tools (e.g., rqt, rviz2) are all based on ROS 2. The complex structure and dependencies of Autoware introduce significant challenges both to human developers and automated testing tools (such as AFL \cite{AFL} and KLEE \cite{cadar2008klee}), in writing correct and meaningful unit tests.

Our study covers both developer-written official test cases and automatically generated test cases. For developer-written test cases, we collect coverage data from all test cases in the Autoware repository, and analyze the reason why these test cases failed to cover certain code elements.
To investigate automatic unit testing techniques,
we crafted an evaluation benchmark from the Autoware source code, which consists of 1126 functions across 8 Autoware software modules.
Due to the difficulties in applying tools for Autoware unit testing, we shift our attention to large language models (LLM), which have demonstrated state-of-the-art results on unit testing for Java/Python software \cite{yuan2024evaluating, wang2024testeval, jain2024testgeneval, ryan2024code, wang2024hits, mundler2024swt}. However, as there are significant differences between Autoware and the datasets adopted in previous LLM-based unit testing, it is questionable whether LLMs can generate useful test cases for Autoware packages.

In summary, we aim to investigate unit testing on Autoware by answering the following research questions:

\begin{compactitem}[$\bullet$]
    \item \textbf{RQ1}: How is the quality of human-written test cases for Autoware packages?

    \item \textbf{RQ2}: How do LLMs perform on unit testing for Autoware?

    \item \textbf{RQ3}: How can we improve the performances of LLMs on Autoware testing?
\end{compactitem}

The results of our study show that for developer-written test cases, their overall test coverage is inadequate: most Autoware packages are left untested, while in packages with test cases, the majority of functions remain uncovered. 
For LLM-generated test cases, we find that LLMs with naive prompt settings perform poorly on Autoware packages. For example, in the \texttt{planning} module, the build success rate of tests generated by GPT-4o-mini is lower than 10\%. To mitigate the errors in LLM-driven Autoware testing, we conduct an empirical analysis of the error types in LLM-generated test cases. Based on the findings in the study, we propose a new approach, AwTest-LLM (\textbf{A}uto\textbf{w}are \textbf{Test}ing with \textbf{LLM}), for Autoware unit testing, incorporating dependency and example extraction from Autoware packages. Experimental results show that our new approach improves both the build success rate and test coverage of LLM-generated test cases for Autoware. 

\section{Study on Human-written Tests}
\label{sec:human}

To assess the state of unit testing in \autoware, we first perform a quantitative and qualitative analysis of the quality of the officially-provided (human-written) test cases.

To execute the human-written test cases, we follow the Autoware documents and utilize \texttt{colcon}, a multi-package ROS build tool supported by Autoware for running unit tests. For our study, we select the \texttt{awsim-stable} branch of the Autoware software repository, as this branch is stable and designed to integrate seamlessly with AWSIM \cite{AWSIM}, an ADS simulator adopted by TIER IV.

For quantitative analysis, we run all official test cases and evaluate their coverage at the package, function, and line levels.
Table~\ref{tab:official} shows the coverage data of official human-written test cases in Autoware. 
We find that 8 out of 10 Autoware software modules are equipped with runnable test cases written by developers. 
The exceptions are the \texttt{perception} and \texttt{sensing} modules, because their performances are more related to deep learning models and sensors than their intrinsic code logic. 
For the 8 modules with test cases, we observe that most packages lack test cases, except for the \texttt{evaluator} and \texttt{simulator} modules. 
This highlights the inadequacy of the current human-written test cases for Autoware. Furthermore, even within packages that have test cases, only a small fraction of functions are covered. 
When focusing on non-trivial functions containing branches, we find that their function coverages (see $funccov_{branch}$) are usually higher than the overall function coverages, in 6 modules out of 8. 
Despite the low overall coverage, if we only consider the covered function, their line coverages (see $linecov_{branch}$) are usually high: all 8 modules have line coverages over 75\% in covered functions with branches.

\begin{table*}
  \caption{Statistics on the coverage of official test cases in Autoware. $pkgcov$, $funccov$, and $linecov$ denotes coverage rate on package/function/line levels, respectively.}
  \label{tab:official}
  \scalebox{0.75}{
  \begin{tabular}{lccccccccccc}
    \toprule
    Module & all pkgs & covered pkgs & $pkgcov$(\%) & \makecell[c]{funcs in \\covered pkgs} & funcs covered & $funccov$(\%) & \makecell[c]{branched funcs \\in covered pkgs} & \makecell[c]{branched func \\covered}& $funccov_{branch}$(\%) & $linecov$(\%)& $linecov_{branch}$(\%)\\
    \midrule
    common & 44 & 14 & 31.8 & 451 & 296 & 65.6 & 169 & 123 & 72.8 & 61.5 & 94.6 \\
    control & 13 & 5 & 38.5 & 204 & 149 & 73.0 & 97 & 78 & 80.4 & 64.5 & 85.6 \\
    evaluator & 4 & 4 & 100.0 & 49 & 46 & 94.0 & 27 & 26 & 96.3 & 86.4 & 93.8\\
    localization & 9 & 3 & 33.3 & 56 & 23 & 41.1 & 17 & 6 & 35.3 & 19.5& 77.8\\
    map	& 4 & 2 & 50.0 & 29 & 5 & 17.2 & 14 & 4 & 28.6 & 11.7 & 86.0 \\
    planning & 21 & 8 & 38.1 & 1313 & 210 & 16.0 & 766 & 125 & 16.3 & 7.3 & 90.8 \\
    simulator & 3 & 3 & 100.0 & 74 & 53 & 71.6 & 23 & 15	& 65.2 & 75.5 & 91.4 \\
    vehicle	& 5 & 1 & 20.0 & 30 & 21 & 70.0 & 16 & 13 & 81.3 & 47.2 & 94.7 \\
    \midrule
    total & 103 & 40 & 38.8 & 2206 & 803 & 36.4 & 1129 & 390 & 34.5 & 17.7 & 89.4 \\
    \bottomrule
  \end{tabular}
  }
\end{table*}



We further perform qualitative analysis on the code not covered by developer-written test cases. Two authors manually categorize the uncovered source code into four types, which are illustrated below.

\begin{compactitem}[$\bullet$]

    \item \textbf{Absence of testing whole files.} 
    We find that many files are not tested by the developers.     
    For example, an important file $ekf\_localizer.cpp$ in the \texttt{localization} module, is used to estimate robot pose by integrating the 2D vehicle dynamics model.
    However, the file is completely untouched by the unit tests, which might hinder the safety of the ADS.
    
    \item \textbf{Absence of testing auxiliary functions.}
    Auxiliary functions are defined to help the operation of the module.
    We discover that many auxiliary functions are not covered by the unit test. For instance, $motion\_velocity\_smoother\_node.cpp$ are designed to plan a velocity profile within the limitations of the velocity, the acceleration and the jerk to realize both the maximization of velocity and the ride quality.
    Function $\texttt{calcTrajectoryVelocity}$ is used to compute the velocity of a given trajectory and is used in $motion\_velocity\_smoother$. 
    However, the function is not tested by the official test cases.
    \begin{lstlisting}
TrajectoryPoints MotionVelocitySmootherNode::calcTrajectoryVelocity(const TrajectoryPoints & traj_input) const { 
... }
\end{lstlisting}
    

    \item \textbf{Absence of testing statements within if-condition.}
    Unit tests cannot test code under an \texttt{if} condition if the condition's logic cannot be satisfied or simulated within the test environment, leading to untested execution paths.
    An example in $gyro\_odometer\_core.cpp$ of \texttt{localization} module, is on checking the timeout using if-condition.
    However, the case of timeout is not triggered by the test cases.

\begin{lstlisting}
if (imu_dt > message_timeout_sec_) {
const std::string error_msg = fmt::format(
"Imu msg is timeout. twist_dt: {}[sec], tolerance {}[sec]", imu_dt, message_timeout_sec_);
...}
\end{lstlisting}

    \item \textbf{Absence of testing exception catching.}
    The $\texttt{catch}$ statement in C++ is used to handle exceptions thrown by the $\texttt{try}$ block, allowing the program to catch specific types of exceptions and define how to handle them.
    An example is a node creation statement in $node\_main\_vehicle\_cmd\_gate.cpp$, where library loading exception is catched.
    Nevertheless, the corresponding test cases do not test this part of code.
    \begin{lstlisting}
try {
node_factory = loader->createInstance<rclcpp_components::NodeFactory>(clazz);} 
catch (const std::exception & ex) {
RCLCPP_ERROR(logger, "Failed to load library %s", ex.what());
return 1;}
\end{lstlisting}

    
\end{compactitem}

\begin{tcolorbox}[size=title,breakable]
\textbf{Answer to RQ1:} \textcolor{black}{The overall test coverage of human-written test cases on Autoware packages is low. The majority of packages are not tested by test cases. In some packages with test cases, most functions are left untested.}
\end{tcolorbox}

\section{Study on Automatic Test Case Generation}
From the empirical study results in Section~\ref{sec:human}, it is clear that the current human-written test cases are insufficient for comprehensive testing on Autoware, so automatically generating test cases can serve as a helpful complement. 
Nevertheless, classic C++ automatic test generation tools, such as AFL~\cite{AFL} and KLEE~\cite{cadar2008klee}, are difficult to be applied in testing \autoware, because of the unsupported multi-processor library, e.g., OpenMP (\autoware depends on), and the difficulty to built the system as LLVM bitcode (both required by AFL and KLEE).

To mitigate this threat, we turn our sight to large language model (LLM), which is a powerful tool for automatic unit test generation~\cite{wang2024hits, lemieux2023codamosa}, free from concerns related to library and compilation.
While LLMs have shown promising results in generating test cases for common software repositories~\cite{yuan2024evaluating, ryan2024code}, their effectiveness on industrial-level software, especially ADS software, is unclear. 


For example, the Autoware system introduces several new challenges in comparison to previous works:

\begin{compactitem}[$\bullet$]
    \item The source code of Autoware is written in C++, which introduces unique language features such as header files, namespaces, and templates. In contrast, most previous LLM for unit testing works focus on Java and Python \cite{yuan2024evaluating, ryan2024code}.
    \item The Autoware packages have complex inter-package dependencies: one Autoware package may depend on multiple other Autoware packages, or even non-Autoware ROS packages. 
    The inter-package dependency information is stored in CMake files.
    \item The inputs and outputs of different modules in ADS are highly diverse. In addition, the modules interact with each other using a messaging mechanism, making the invocation process more complex.
\end{compactitem}

To measure the effectiveness of LLMs on Autoware unit testing, we create a benchmark dataset based on our study results in Section~\ref{sec:human}. 
From all functions in packages being tested by developers, we collect functions with branches and input arguments. 
As a result, we create two datasets for evaluation: \textbf{the covered dataset:} 390 functions covered by official test cases, and \textbf{the uncovered dataset:} 812 functions not covered by official test cases.

In the remaining subsections, we show an early-step exploration on using LLMs in unit test generation for ADS software.
 



\subsection{Experiment Setup}

We run the test case generation and evaluation pipeline in the following steps:

\begin{compactitem}[$\bullet$]
    \item 1. We ask the LLM to generate test cases for all functions under test given the function source code and corresponding contexts. For each function under test, we prompt the LLM to generate a single test file with one or more test cases in order to maximize code coverage.

    \item 2. We modify the \texttt{CmakeLists.txt} file in all packages to add the generated tests to the package.

    \item 3. We use \texttt{colcon} commands to build and run all test cases.

    \item 4. We use \texttt{lcov}~\footnote{\texttt{lcov} is run through its \texttt{colcon} extension: https://github.com/colcon/colcon-lcov-result} to measure the coverage of test cases.
\end{compactitem}

We adopt two LLMs: GPT-4o-mini and GPT-4o for our experiments. The generated test cases are measured by 3 metrics: build success rate ($BS$), run success rate ($RS$), and line coverage. We do not record branch coverage, since the "branch coverage" computed by \texttt{lcov} is inconsistent with the branches in the source code. $BS$ and $RS$ are first computed on test file level. Because one test file may contain multiple test cases, we further measure $RS$ on test case level ($RS_{case}$), which is computed by: 

\begin{equation}
    RS_{case} = \frac{No.\ test\ cases\ without\ runtime\ errors}{No.\ test\ cases\ in\ successfully\ built\ test\ files}
\end{equation}

We first run test case generation with a basic prompt setting: the LLM is given the function signature and the whole C++ file of the focal function. We use the whole focal file as the input because it may contain useful context information, such as the definition of functions called within the focal function. We have also run experiments that only give LLMs the focal function instead of the complete file, and their build success rates are lower than our basic setting.

\begin{table*}[htbp]
  \caption{Results on the covered dataset with the basic prompt setting.}
  \label{tab:result-basic}
  \scalebox{0.8}{
  \begin{tabular}{lccccccccccc}
    \toprule
    \multirow{2}{*}{Module} & \multirow{2}{*}{functions} & \multicolumn{2}{c}{$BS_{file}(\%)$} & \multicolumn{2}{c}{$RS_{file}(\%)$} & \multicolumn{2}{c}{built test cases} & \multicolumn{2}{c}{$RS_{case}(\%)$} & \multicolumn{2}{c}{line coverage(\%)}\\
    \cmidrule(lr){3-4} \cmidrule{5-6} \cmidrule(lr){7-8} \cmidrule(l){9-10} \cmidrule(l){11-12}
    & & GPT-4o-mini & GPT-4o & GPT-4o-mini & GPT-4o & GPT-4o-mini & GPT-4o & GPT-4o-mini & GPT-4o & GPT-4o-mini & GPT-4o \\
    \midrule
    common & 123 & \textbf{41.5} & 37.4 & \textbf{38.2} & 35.8 & 331 & \textbf{405} & 73.7 & \textbf{74.1} & \textbf{30.3} & 16.9 \\
    control & 78 & 14.1 & \textbf{30.8} & 12.8 & \textbf{21.8} & 65 & \textbf{186} & 55.4 & \textbf{59.1} & 5.7 & \textbf{9.5}\\
    evaluator & 26 & 15.4 & 15.4 & 11.5 & \textbf{15.4} & 21 & \textbf{37} & 71.4 & \textbf{86.5} & 2.7 & \textbf{7.6}\\
    localization & 6 & 16.7 & 16.7 & 16.7 & 16.7 & 9 & 9 & \textbf{100.0} & 88.9 & 3.4 & 3.4\\
    map & 4 & 50.0 & 50.0 & 50.0 & 50.0 & 10 & 10 & \textbf{50.0} & 40.0 & 22.8 & 22.8 \\
    planning & 125 & 9.6 & \textbf{12.8} & 8.8 & \textbf{12.8} & 74 & \textbf{212} & 66.2 & \textbf{75.0} & \textbf{5.3} & 4.9\\
    simulator & 15 & 13.3 & \textbf{20.0} & 13.3 & \textbf{20.0} & 11 & 16 & \textbf{72.7} & 31.3 & 3.1 & \textbf{5.8} \\
    vehicle & 13 & 30.8 & \textbf{61.5} & 7.7 & \textbf{53.8} & 14 & \textbf{79} & 50.0 & \textbf{81.0} & 19.1 & \textbf{53.4}\\
    \midrule
    overall & 390 & 22.3 & \textbf{26.7} & 19.7 & \textbf{24.1} & 535 & \textbf{954} & 69.7 & \textbf{71.5} & \textbf{13.9} & 11.4 \\
    \bottomrule
  \end{tabular}
    }
\end{table*}

\subsection{Results and Analysis: the Basic Setting}

Table~\ref{tab:result-basic} demonstrates the success rates and coverage for LLMs using the basic prompt setting. The results of LLMs are far from satisfactory compared to the results on Java/Python repositories \cite{yuan2024evaluating, ryan2024code}. 
For GPT-4o-mini, \textbf{none} of the modules achieved build pass rates over 50\%. Apart from \texttt{common}, \texttt{map}, and \texttt{vehicle} modules, the $BS_{file}$ on other packages are lower than 20\%. Line coverage was similarly low, with 5 modules achieving less than 10\%, primarily due to the low build pass rates. 
The overall $BS_{file}$ and $RS_{file}$ of GPT-4o are marginally higher than GPT-4o-mini, but still below 30\%. 
The gaps between $BS_{file}$ and $RS_{file}$ for most modules are small, indicating that for each test file that can be built, there exists at least one test case that can run without error. 
This is further strengthened by the results on $RS_{case}$, where both models achieve case-level correctness around 70\%. In terms of test coverage, the results are low: in 5 out of 8 modules, both LLMs have line coverage lower than 10\%.

From the above results, we can conclude that \textbf{building errors} are the main obstacle that prevents LLM-based testing from generating useful test cases. 
To better understand these errors, we conduct a study to identify the root causes of the test case failures. Specifically, we collect all build reports generated by CMake for failed test cases and perform a manual analysis to determine the underlying causes. 
We perform manual analysis on build errors because the actual root cause of an error may be different from the one described in the build report. 
For instance, a `function not declared' error may be caused by failing to include the header file that contains the function.

\begin{table}[htbp]
  \caption{Analysis results on build error types of test cases generated by GPT-4o-mini with the basic prompt setting.}
  \label{tab:builderror}
  \scalebox{0.8}{
  \begin{tabular}{llc}
    \toprule
    Category & Detailed Error Type & Count \\
    \midrule
    \multirow{2}{*}{Namespace error} & Missing namespace & 37 \\
    & incorrect namespace & 34\\
    \midrule
    \multirow{3}{*}{Symbol error} & Invoking nonexistent member & 128\\ 
    & Misuse existing member & 26\\
    & Use before definition & 3\\
    \midrule
    Type error & Type inconsistency & 61 \\
    \midrule
    \multirow{3}{*}{Header error} & Include nonexistent headers & 74\\
    & Missing necessary headers & 8 \\
    & function/class not in headers & 8 \\
    \midrule
    Syntax error & -- & 10 \\
    \midrule
    \multirow{2}{*}{Access error} & Private access & 97 \\
    & Protected access & 11 \\
    \midrule
    Other error & -- & 30 \\
    \bottomrule
  \end{tabular}
  }
\end{table}

The results of our manual analysis on build errors are shown in Table~\ref{tab:builderror}. 
The most prevalent error type is symbol error ($29.8\%$), which means the LLM mistakenly uses nonexistent functions/classes or misuses existing ones. 
This suggests that LLMs are prone to hallucinations when generating test cases, and cannot fully understand some functions/classes in Autoware packages. Other common error types include incorrectly calling private methods, type inconsistencies, and namespace errors.

\begin{table}[htbp]
  \caption{Breakdown of runtime error types in test cases generated by GPT-4o-mini with the basic prompt setting.}
  \label{tab:runerror}
  \scalebox{0.8}{
  \begin{tabular}{llc}
    \toprule
    Category & Detailed Error Type & Count \\
    \midrule
    \multirow{3}{*}{Assertion error} & Assertion on values & \multirow{3}{*}{134} \\
    & Assertion on throw & \\
    & Assertion on death behaviors & \\
    \midrule
    \multirow{2}{*}{Runtime exceptions} & Out of range & \multirow{2}{*}{28} \\
    & Invalid argument & \\
    \midrule
    \multirow{2}{*}{Other errors} & \multirow{2}{*}{Timeout} & \multirow{2}{*}{5} \\ & \\
    \bottomrule
  \end{tabular}
  }
\end{table}

We further analyze the runtime error types in LLM-generated test cases, which its results are shown in Table~\ref{tab:runerror}. As demonstrated in the table, assertion error is the main type of runtime error, suggesting that LLMs struggle to accurately capture the expected behavior of functions in Autoware. Other runtime error types include some typical C++ exceptions, such as out-of-range and invalid arguments.

\begin{tcolorbox}[size=title,breakable]
\textbf{Answer to RQ2:} LLMs exhibit low pass rates and test coverage in unit testing for Autoware. The main reason for the failure of LLM-generated test cases is the hallucinations in invoking incorrect functions, and difficulties in understanding C++-specific grammar patterns.
\end{tcolorbox}

\subsection{Improving LLM-based Test Case Generation}

\subsubsection{Proposed approach}

From the analysis results on LLM-generated test cases, we can see that building errors are still the most critical challenge in LLM-driven testing for Autoware. 
To mitigate this threat, we propose a new LLM-based approach, AwTest-LLM, to generate unit test cases for Autoware packages.
Figure~\ref{fig:awtest} shows the overall pipeline of our proposed approach. AwTest-LLM addresses the challenges in generating correct Autoware test cases with the following steps:

\begin{figure}[h]
  \centering
  \includegraphics[width=0.9\linewidth]{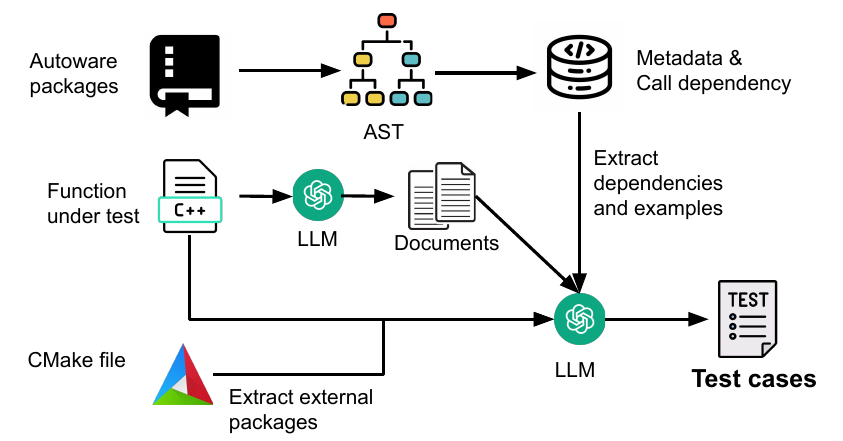}
  \caption{The overview of AwTest-LLM.}
  \Description{}
  \label{fig:awtest}
\end{figure}

\begin{enumerate}
    \item \textbf{Packages Preprocessing}: For all Autoware packages, we analyze all C++ files with an abstract syntax tree (AST) parser. From the generated ASTs, we extract dependency information for each function, including its namespace, associated header files, and class metadata(if the function is a class method). 
    
    \item \textbf{Call Dependency Extraction}: We also build a call graph from ASTs for each package to extract call dependencies between functions. From the call graph, we can extract functions that call the focal function (i.e., function under test). These calling functions can serve as few-shot in-context-learning \cite{brown2020language} examples, helping LLMs to correctly generate test cases by demonstrating appropriate usage of the focal function.

    \item \textbf{Document Generation}: In Autoware, some functions have comprehensive documents written by developers, while others do not. For focal functions without corresponding documentation, we prompt the LLM to generate a description that includes the function's functionality, input arguments, and return values.

    \item \textbf{CMake File Parsing}: Some Autoware packages rely on external Autoware/ROS packages. We parse the CMake file in each package to extract the external package dependency information.

    \item \textbf{Test Case Generation}: With all metadata and dependencies extracted, we construct the final test case generation prompt using the focal file, along with the namespace, headers, function document, external dependencies, and examples.
\end{enumerate}

When extracting metadata and dependencies, we adopt tree-sitter \cite{tree-sitter} to parse C++ code into ASTs. 
For a few functions that cannot be correctly parsed to ASTs due to tree-sitter bugs, we return to the basic prompt for test case generation.

\subsubsection{Experimental results}

\begin{table*}[ht]
  \caption{Results on the covered dataset with AwTest-LLM. The results in brackets are the improvements over the basic prompt setting.}
  \vspace{-10pt}
  \label{tab:result-awtest}
  \scalebox{0.8}{
  \begin{tabular}{lccccccccccc}
    \toprule
    \multirow{2}{*}{Module} & \multirow{2}{*}{functions} & \multicolumn{2}{c}{$BS_{file}$(\%)} & \multicolumn{2}{c}{$RS_{file}(\%)$} & \multicolumn{2}{c}{built test cases} & \multicolumn{2}{c}{$RS_{case}(\%)$} & \multicolumn{2}{c}{line coverage(\%)}\\
    \cmidrule(lr){3-4} \cmidrule{5-6} \cmidrule(lr){7-8} \cmidrule(l){9-10} \cmidrule(l){11-12}
    & & GPT-4o-mini & GPT-4o & GPT-4o-mini & GPT-4o & GPT-4o-mini & GPT-4o & GPT-4o-mini & GPT-4o & GPT-4o-mini & GPT-4o \\
    \midrule
    common & 123 & 44.7(+3.2) & \textbf{56.9(+19.5)} & 39.8(+1.6) & \textbf{53.7(+17.9)} & 322 & 339 & 73.9(+0.2) & \textbf{79.6(+5.5)} & 32.8(+2.5) & \textbf{45.9(+29.0)} \\
    control & 78 & \textbf{30.8(+16.7)} & 19.2(-11.6) & \textbf{26.9(+14.1)} & 19.2(-2.6) & 118 & 86 & 56.8(+1.4) & \textbf{58.1(-1.0)} & \textbf{13.0(+7.3)} & 9.6(+0.1) \\
    evaluator & 26 & \textbf{19.2(+3.8)} & 15.4(+0) & \textbf{15.4(+3.8)} & 15.4(+0) & 16 & 22 & 18.8(-53.6) & \textbf{59.1(-26.4)} & \textbf{7.2(+4.5)} & 2.7(-4.9) \\
    localization & 6 & 16.7(+0) & 16.7(+0) & 16.7(+0) & 16.7(+0) & 9 & 5 & 100.0(+0) & 100.0(+0) & 3.4(+0) & 3.4(+0)\\
    map & 4 & 50.0(+0) & 50.0(+0) & 50.0(+0) & 50.0(+0) & 10 & 9 & 50.0(+0) & \textbf{55.6(+15.6)} & 22.8(+0) & \textbf{24.6(+1.8)}\\
    planning & 125 & 11.2(+1.6) & \textbf{20.8(+8.0)} & 10.4(+1.6) & \textbf{17.6(+4.8)} & 75 & 111 & 52.0(-14.2) & \textbf{62.2(-12.8)} & 9.0(+3.7) & \textbf{13.1(+8.2)} \\
    simulator & 15 & 20.0(+6.7) & \textbf{26.7(+6.7)} & 20.0(+6.7) & 20.0(+0) & 14 & 18 & \textbf{78.6(+5.9)} & 72.2(+40.9) & \textbf{9.3(+6.2)} & 7.1(+1.3) \\
    vehicle & 13 & \textbf{84.6(+53.8)} & 61.5(+0) & 38.5(+30.8) & \textbf{46.2(-7.6)} & 49 & 37 & 85.7(+35.7) & \textbf{89.2(+8.2)} & 38.9(+19.8) & \textbf{45.8(-7.6)} \\
    \midrule
    overall & 390 & 29.5(+7.2) & \textbf{33.3(+6.6)} & 25.1(+5.4) & \textbf{30.5(+6.4)} & 613 & 627 & 67.5(-2.2) & \textbf{73.0(+1.5)} & 18.4(+4.5) & \textbf{23.0(+11.6)} \\
    \bottomrule
  \end{tabular}
    }
    \vspace{-10pt}
\end{table*}

\begin{table*}[ht]
  \caption{Results on the uncovered dataset with AwTest-LLM. }
 \vspace{-10pt}
  \label{tab:result-uncovered}
  \scalebox{0.8}{
  \begin{tabular}{lccccccccccc}
    \toprule
    \multirow{2}{*}{Module} & \multirow{2}{*}{functions} & \multicolumn{2}{c}{$BS_{file}(\%)$} & \multicolumn{2}{c}{$RS_{file}(\%)$} & \multicolumn{2}{c}{built test cases} & \multicolumn{2}{c}{$RS_{case}(\%)$} & \multicolumn{2}{c}{line coverage(\%)}\\
    \cmidrule(lr){3-4} \cmidrule{5-6} \cmidrule(lr){7-8} \cmidrule(l){9-10} \cmidrule(l){11-12}
    & & GPT-4o-mini & GPT-4o & GPT-4o-mini & GPT-4o & GPT-4o-mini & GPT-4o & GPT-4o-mini & GPT-4o & GPT-4o-mini & GPT-4o \\
    \midrule
    common & 46 & 30.4 & \textbf{37.0} & 30.4 & \textbf{34.8} & 82 & 86 & 63.4 & \textbf{70.9} & \textbf{36.2} & 30.2 \\
    control & 19 & 0.0 & \textbf{5.3} & 0.0 & \textbf{5.3} & 0 & 5 & 0 & \textbf{100.0} & 0 & 0.0 \\
    evaluator & 1 & 0.0 & 0.0 & 0.0 & 0.0 & 0 & 0 & 0 & 0 & 0 & 0 \\
    localization & 10& 10.0 & 10.0 & 10.0 & 10.0 & 7 & 4 & 85.7 & \textbf{100.0} & 6.2 & 6.2 \\
    map & 10 & 0.0 & \textbf{10.0} & 0.0 & \textbf{10.0} & 0 & 6 & 0 & \textbf{100.0} & 0 & 0.0 \\
    planning & 639 & 10.3 & \textbf{14.7} & 8.8 & \textbf{13.1} & 299 & 311 & 56.2 & \textbf{68.8} & 4.2 & \textbf{6.0}\\
    simulator & 8 & 0.0 & 0.0 & 0.0 & 0.0 & 0 & 0 & 0 & 0 & 0 & 0 \\
    vehicle & 3 & 33.3 & \textbf{100.0} & 33.3 & \textbf{100.0} & 5 & 15 & 0 & \textbf{20.0} & 0 & \textbf{24.4} \\
    \midrule
    overall & 736 & 11.5 & \textbf{15.9} & 10.1 & \textbf{14.4} & 393 & 427 & 57.5 & \textbf{68.6} & 5.2 & \textbf{6.7} \\
    \bottomrule
  \end{tabular}
    }
\end{table*}

Table~\ref{tab:result-awtest} shows the results of AwTest-LLM on the covered dataset.
We find that AwTest-LLM brings improvements in overall $BS$ and $RS$. Notably, $BS_{file}$ and $RS_{file}$ both increase in 6 out of 8 modules for GPT-4o-mini. For the \texttt{control} and \texttt{vehicle} modules, their improvements are over 15\%. For the modules with the most focal functions, i.e., \texttt{common} and \texttt{planning}, the improvements of AwTest-LLM on both LLMs are also prominent. Along with increased success rates, we can also see improvements in test coverage: the overall line coverage has raised 11.6\% for GPT-4o. However, since AwTest-LLM does not specifically address the generation of correct test assertions, there are no significant improvements on $RS_{case}$ over the basic prompt setting.

After we witnessed the improvements of AwTest-LLM on the covered dataset, we further evaluate its performance on the uncovered dataset, as shown in Table~\ref{tab:result-uncovered}. LLMs exhibit lower pass rates and coverage on the uncovered dataset, indicating that functions not previously tested by human developers are generally more challenging to test. Except for the \texttt{common} and \texttt{vehicle} modules, the coverages of AwTest-LLM with GPT-4o on other modules are all lower than 10\%, which suggests that it is still very challenging to test complete Autoware packages with LLMs.

\begin{tcolorbox}[size=title,breakable]
\textbf{Answer to RQ3:} Experiment results suggest that AwTest-LLM enhances the pass rates and coverage of LLM-generated test cases for Autoware, demonstrating the potential of applying LLM-driven testing into the real-world development process. However, significant challenges remain in generating test cases for functions that have not been previously tested by human developers.
\end{tcolorbox}

\section{Conclusion}
In this paper, we conduct an early-stage study on fine-grained unit testing of autonomous driving systems. 
Centered on Autoware, we studied the current state of developer-written test cases, and how LLMs performed in automatic test case generation.
Based on our findings, we proposed AwTest-LLM, a novel LLM-based unit testing framework for Autoware. Compared to the basic usage of LLMs, AwTest-LLM can boost the build success rate and code coverage for test cases generated from Autoware packages. 
In the future, we aim to continuously improve the Awtest-LLM framework to achieve higher pass rates and improved coverage for generated test cases.


\bibliographystyle{ACM-Reference-Format}
\bibliography{sample-base}


\begin{thebibliography}{19}


\ifx \showCODEN    \undefined \def \showCODEN     #1{\unskip}     \fi
\ifx \showDOI      \undefined \def \showDOI       #1{#1}\fi
\ifx \showISBNx    \undefined \def \showISBNx     #1{\unskip}     \fi
\ifx \showISBNxiii \undefined \def \showISBNxiii  #1{\unskip}     \fi
\ifx \showISSN     \undefined \def \showISSN      #1{\unskip}     \fi
\ifx \showLCCN     \undefined \def \showLCCN      #1{\unskip}     \fi
\ifx \shownote     \undefined \def \shownote      #1{#1}          \fi
\ifx \showarticletitle \undefined \def \showarticletitle #1{#1}   \fi
\ifx \showURL      \undefined \def \showURL       {\relax}        \fi
\providecommand\bibfield[2]{#2}
\providecommand\bibinfo[2]{#2}
\providecommand\natexlab[1]{#1}
\providecommand\showeprint[2][]{arXiv:#2}

\bibitem[AFL(2013)]%
        {AFL}
 \bibinfo{year}{2013}\natexlab{}.
\newblock \bibinfo{booktitle}{\emph{american fuzzy lop}}.
\newblock
\urldef\tempurl%
\url{https://github.com/google/AFL}
\showURL{%
\tempurl}


\bibitem[tre(2019)]%
        {tree-sitter}
 \bibinfo{year}{2019}\natexlab{}.
\newblock \bibinfo{booktitle}{\emph{Tree-sitter}}.
\newblock
\urldef\tempurl%
\url{https://tree-sitter.github.io/tree-sitter/}
\showURL{%
\tempurl}


\bibitem[ROS(2022)]%
        {ROS}
 \bibinfo{year}{2022}\natexlab{}.
\newblock \bibinfo{booktitle}{\emph{ROS - Robot Operating System}}.
\newblock
\urldef\tempurl%
\url{https://www.ros.org/}
\showURL{%
\tempurl}


\bibitem[aut(2024)]%
        {autoware}
 \bibinfo{year}{2024}\natexlab{}.
\newblock \bibinfo{booktitle}{\emph{Autoware - the world's leading open-source software project for autonomous driving}}.
\newblock
\urldef\tempurl%
\url{https://github.com/autowarefoundation/autoware}
\showURL{%
\tempurl}


\bibitem[AWS(2024)]%
        {AWSIM}
 \bibinfo{year}{2024}\natexlab{}.
\newblock \bibinfo{booktitle}{\emph{AWSIM}}.
\newblock
\urldef\tempurl%
\url{https://github.com/tier4/AWSIM}
\showURL{%
\tempurl}


\bibitem[Brown et~al\mbox{.}(2020)]%
        {brown2020language}
\bibfield{author}{\bibinfo{person}{Tom Brown}, \bibinfo{person}{Benjamin Mann}, \bibinfo{person}{Nick Ryder}, \bibinfo{person}{Melanie Subbiah}, \bibinfo{person}{Jared~D Kaplan}, \bibinfo{person}{Prafulla Dhariwal}, \bibinfo{person}{Arvind Neelakantan}, \bibinfo{person}{Pranav Shyam}, \bibinfo{person}{Girish Sastry}, \bibinfo{person}{Amanda Askell}, {et~al\mbox{.}}} \bibinfo{year}{2020}\natexlab{}.
\newblock \showarticletitle{Language models are few-shot learners}.
\newblock \bibinfo{journal}{\emph{Advances in neural information processing systems}}  \bibinfo{volume}{33} (\bibinfo{year}{2020}), \bibinfo{pages}{1877--1901}.
\newblock


\bibitem[Cadar et~al\mbox{.}(2008)]%
        {cadar2008klee}
\bibfield{author}{\bibinfo{person}{Cristian Cadar}, \bibinfo{person}{Daniel Dunbar}, \bibinfo{person}{Dawson~R Engler}, {et~al\mbox{.}}} \bibinfo{year}{2008}\natexlab{}.
\newblock \showarticletitle{Klee: unassisted and automatic generation of high-coverage tests for complex systems programs.}. In \bibinfo{booktitle}{\emph{OSDI}}, Vol.~\bibinfo{volume}{8}. \bibinfo{pages}{209--224}.
\newblock


\bibitem[Cheng et~al\mbox{.}(2023)]%
        {cheng2023behavexplor}
\bibfield{author}{\bibinfo{person}{Mingfei Cheng}, \bibinfo{person}{Yuan Zhou}, {and} \bibinfo{person}{Xiaofei Xie}.} \bibinfo{year}{2023}\natexlab{}.
\newblock \showarticletitle{Behavexplor: Behavior diversity guided testing for autonomous driving systems}. In \bibinfo{booktitle}{\emph{Proceedings of the 32nd ACM SIGSOFT International Symposium on Software Testing and Analysis}}. \bibinfo{pages}{488--500}.
\newblock


\bibitem[Jain et~al\mbox{.}(2024)]%
        {jain2024testgeneval}
\bibfield{author}{\bibinfo{person}{Kush Jain}, \bibinfo{person}{Gabriel Synnaeve}, {and} \bibinfo{person}{Baptiste Rozi{\`e}re}.} \bibinfo{year}{2024}\natexlab{}.
\newblock \showarticletitle{Testgeneval: A real world unit test generation and test completion benchmark}.
\newblock \bibinfo{journal}{\emph{arXiv preprint arXiv:2410.00752}} (\bibinfo{year}{2024}).
\newblock


\bibitem[Lemieux et~al\mbox{.}(2023)]%
        {lemieux2023codamosa}
\bibfield{author}{\bibinfo{person}{Caroline Lemieux}, \bibinfo{person}{Jeevana~Priya Inala}, \bibinfo{person}{Shuvendu~K Lahiri}, {and} \bibinfo{person}{Siddhartha Sen}.} \bibinfo{year}{2023}\natexlab{}.
\newblock \showarticletitle{Codamosa: Escaping coverage plateaus in test generation with pre-trained large language models}. In \bibinfo{booktitle}{\emph{2023 IEEE/ACM 45th International Conference on Software Engineering (ICSE)}}. IEEE, \bibinfo{pages}{919--931}.
\newblock


\bibitem[Lou et~al\mbox{.}(2022)]%
        {lou2022testing}
\bibfield{author}{\bibinfo{person}{Guannan Lou}, \bibinfo{person}{Yao Deng}, \bibinfo{person}{Xi Zheng}, \bibinfo{person}{Mengshi Zhang}, {and} \bibinfo{person}{Tianyi Zhang}.} \bibinfo{year}{2022}\natexlab{}.
\newblock \showarticletitle{Testing of autonomous driving systems: where are we and where should we go?}. In \bibinfo{booktitle}{\emph{Proceedings of the 30th ACM Joint European Software Engineering Conference and Symposium on the Foundations of Software Engineering}}. \bibinfo{pages}{31--43}.
\newblock


\bibitem[M{\"u}ndler et~al\mbox{.}(2024)]%
        {mundler2024swt}
\bibfield{author}{\bibinfo{person}{Niels M{\"u}ndler}, \bibinfo{person}{Mark~Niklas Mueller}, \bibinfo{person}{Jingxuan He}, {and} \bibinfo{person}{Martin Vechev}.} \bibinfo{year}{2024}\natexlab{}.
\newblock \showarticletitle{SWT-Bench: Testing and Validating Real-World Bug-Fixes with Code Agents}. In \bibinfo{booktitle}{\emph{The Thirty-eighth Annual Conference on Neural Information Processing Systems}}.
\newblock


\bibitem[Ryan et~al\mbox{.}(2024)]%
        {ryan2024code}
\bibfield{author}{\bibinfo{person}{Gabriel Ryan}, \bibinfo{person}{Siddhartha Jain}, \bibinfo{person}{Mingyue Shang}, \bibinfo{person}{Shiqi Wang}, \bibinfo{person}{Xiaofei Ma}, \bibinfo{person}{Murali~Krishna Ramanathan}, {and} \bibinfo{person}{Baishakhi Ray}.} \bibinfo{year}{2024}\natexlab{}.
\newblock \showarticletitle{Code-aware prompting: A study of coverage-guided test generation in regression setting using llm}.
\newblock \bibinfo{journal}{\emph{Proceedings of the ACM on Software Engineering}} \bibinfo{volume}{1}, \bibinfo{number}{FSE} (\bibinfo{year}{2024}), \bibinfo{pages}{951--971}.
\newblock


\bibitem[Tang et~al\mbox{.}(2023)]%
        {tang2023survey}
\bibfield{author}{\bibinfo{person}{Shuncheng Tang}, \bibinfo{person}{Zhenya Zhang}, \bibinfo{person}{Yi Zhang}, \bibinfo{person}{Jixiang Zhou}, \bibinfo{person}{Yan Guo}, \bibinfo{person}{Shuang Liu}, \bibinfo{person}{Shengjian Guo}, \bibinfo{person}{Yan-Fu Li}, \bibinfo{person}{Lei Ma}, \bibinfo{person}{Yinxing Xue}, {et~al\mbox{.}}} \bibinfo{year}{2023}\natexlab{}.
\newblock \showarticletitle{A survey on automated driving system testing: Landscapes and trends}.
\newblock \bibinfo{journal}{\emph{ACM Transactions on Software Engineering and Methodology}} \bibinfo{volume}{32}, \bibinfo{number}{5} (\bibinfo{year}{2023}), \bibinfo{pages}{1--62}.
\newblock


\bibitem[Tian et~al\mbox{.}(2018)]%
        {tian2018deeptest}
\bibfield{author}{\bibinfo{person}{Yuchi Tian}, \bibinfo{person}{Kexin Pei}, \bibinfo{person}{Suman Jana}, {and} \bibinfo{person}{Baishakhi Ray}.} \bibinfo{year}{2018}\natexlab{}.
\newblock \showarticletitle{Deeptest: Automated testing of deep-neural-network-driven autonomous cars}. In \bibinfo{booktitle}{\emph{Proceedings of the 40th international conference on software engineering}}. \bibinfo{pages}{303--314}.
\newblock


\bibitem[Wang et~al\mbox{.}(2024b)]%
        {wang2024testeval}
\bibfield{author}{\bibinfo{person}{Wenhan Wang}, \bibinfo{person}{Chenyuan Yang}, \bibinfo{person}{Zhijie Wang}, \bibinfo{person}{Yuheng Huang}, \bibinfo{person}{Zhaoyang Chu}, \bibinfo{person}{Da Song}, \bibinfo{person}{Lingming Zhang}, \bibinfo{person}{An~Ran Chen}, {and} \bibinfo{person}{Lei Ma}.} \bibinfo{year}{2024}\natexlab{b}.
\newblock \showarticletitle{TESTEVAL: Benchmarking Large Language Models for Test Case Generation}.
\newblock \bibinfo{journal}{\emph{arXiv preprint arXiv:2406.04531}} (\bibinfo{year}{2024}).
\newblock


\bibitem[Wang et~al\mbox{.}(2024a)]%
        {wang2024hits}
\bibfield{author}{\bibinfo{person}{Zejun Wang}, \bibinfo{person}{Kaibo Liu}, \bibinfo{person}{Ge Li}, {and} \bibinfo{person}{Zhi Jin}.} \bibinfo{year}{2024}\natexlab{a}.
\newblock \showarticletitle{HITS: High-coverage LLM-based Unit Test Generation via Method Slicing}. In \bibinfo{booktitle}{\emph{Proceedings of the 39th IEEE/ACM International Conference on Automated Software Engineering}}. \bibinfo{pages}{1258--1268}.
\newblock


\bibitem[Yuan et~al\mbox{.}(2024)]%
        {yuan2024evaluating}
\bibfield{author}{\bibinfo{person}{Zhiqiang Yuan}, \bibinfo{person}{Mingwei Liu}, \bibinfo{person}{Shiji Ding}, \bibinfo{person}{Kaixin Wang}, \bibinfo{person}{Yixuan Chen}, \bibinfo{person}{Xin Peng}, {and} \bibinfo{person}{Yiling Lou}.} \bibinfo{year}{2024}\natexlab{}.
\newblock \showarticletitle{Evaluating and improving chatgpt for unit test generation}.
\newblock \bibinfo{journal}{\emph{Proceedings of the ACM on Software Engineering}} \bibinfo{volume}{1}, \bibinfo{number}{FSE} (\bibinfo{year}{2024}), \bibinfo{pages}{1703--1726}.
\newblock


\bibitem[Zhang et~al\mbox{.}(2018)]%
        {zhang2018deeproad}
\bibfield{author}{\bibinfo{person}{Mengshi Zhang}, \bibinfo{person}{Yuqun Zhang}, \bibinfo{person}{Lingming Zhang}, \bibinfo{person}{Cong Liu}, {and} \bibinfo{person}{Sarfraz Khurshid}.} \bibinfo{year}{2018}\natexlab{}.
\newblock \showarticletitle{DeepRoad: GAN-based metamorphic testing and input validation framework for autonomous driving systems}. In \bibinfo{booktitle}{\emph{Proceedings of the 33rd ACM/IEEE international conference on automated software engineering}}. \bibinfo{pages}{132--142}.
\newblock


\end{thebibliography}

\appendix

\end{document}